\documentclass[a4paper, 12pt]{article}

\pdfoutput=1

\usepackage{a4}
\usepackage{amsfonts}
\usepackage{amssymb}
\usepackage{cite}
\usepackage[all]{xy}
\usepackage{amsmath}

\usepackage{mathptmx}       
\usepackage{helvet}         
\usepackage{courier}        
\usepackage{type1cm}        
\usepackage{graphicx}        
\usepackage[bottom]{footmisc}

\begin{document}

\title{
{\bf  On the Symmetries of Classical String Theory} }
\vspace{0.1cm}
\author{Constantin Bachas}

\date{}

\maketitle

\begin{center}
\emph{  Laboratoire de Physique Th\'eorique de l'\'Ecole Normale Sup\'erieure, \\
24 rue Lhomond, 75231 Paris,
France \footnote{Unit\'e mixte de recherche (UMR 8549)
du CNRS  et de l'ENS, associ\'ee \`a l'Universit\'e  Pierre et Marie Curie et aux
f\'ed\'erations de recherche
FR684  et FR2687.}}\\ 
\vspace{0.3cm}


\end{center}

\begin{abstract}
\noindent  
\vskip 3mm\ 
I discuss some aspects of conformal defects and conformal
interfaces in two spacetime dimensions. Special emphasis is placed on their
role as spectrum-generating symmetries of classical string theory. 
Contributed  to the  volume 
celebrating Claudio Bunster's sixtieth anniversary;  
based on
 talks at the Arnold Sommerfeld Workshop on ``String Field Theory and Related
Aspects", and for the 50th anniversary of the IHES . 

\end{abstract}

\thispagestyle{empty}
\clearpage

\section{Loop operators in 2d CFT}
\label{sec:1}
   Wilson loops 
  \cite{Wilson:1974sk} are important tools  for the study  of gauge theory.
   They   describe worldlines  of external probes, such as the
  heavy quarks of QCD,  which transform in some representation  of the gauge
  group and  couple to the gauge fields  minimally.  More general couplings, possibly
  involving other  fields  (e.g. scalars and fermions),  are in principle also allowed. They are,
however,  severely limited  by the requirement of infrared relevance  or,  equivalently,  of 
  renormalizability. In four dimensions this  only allows couplings
  to operators of dimension at most one, i.e. 
 linear in the gauge and the scalar fields. 
 An example in which
the scalar coupling  plays  a  role is  the supersymmetric Wilson loop of $N=4$ super-Yang Mills
theory  \cite{Mald, Rey:1998ik}.
 
   The story  is much richer in two space-time dimensions.  
   Power-counting renormalizable defects in a two-dimensional  non-linear sigma model, for example,
   are described by the following loop operators
  \begin{equation}
  \label{1} 
   {\rm tr }_V {W}(C)  =   {\rm tr }_V  \,  {\rm P} \, e^{\, i\oint_{\, C}  \,  {\rm H}_{\rm def} }\ \ ,  
 \end{equation}
where $V$ is the $n$-dimensional space of quantum states of the 
external probe, whose
 Hamiltonian is of the general form 
  \begin{equation} 
  \label{gen1form}
       \oint_C\,   {\rm H}_{\rm def} =   \int  ds   \,  \Bigl [  \left(
        \vec {\mathbf{B}}_M (\Phi)    \partial_\alpha { { \Phi}}^M   +
       \epsilon_{\alpha\beta} \widetilde {\mathbf{B}}_M (\Phi) \partial^\beta \Phi^M \right) 
       {d \hat \zeta^\alpha\over ds} 
       + {\mathbf{T}}(\Phi) \Bigr ]  \ . 
  \end{equation} 
Here $s$ is the length along the defect worldline  $C$, and 
the Hamiltonian  is a hermitean $n\times n$ matrix  which  depends
on the sigma-model fields $\Phi^M(\zeta^{\alpha})$,  and on
 their first derivatives,   evaluated at the position of the defect $\hat \zeta^\alpha(s)$.
  The  loop operator  is thus  specified by  two
  matrix-valued one-forms, 
$ {\mathbf{B}}_M d \Phi^M$ and $\widetilde {\mathbf{B}}_M d\Phi^M $,  and by
a matrix-valued function,  $ {\mathbf{T}}$, all defined 
on  the sigma-model  target space ${\cal M}$.  Because $ {\rm H}_{\rm def} $  is  a matrix,   the 
 path-ordering in (\ref{1}) is non-trivial  even if  the bulk fields are treated as classical,   
 and hence as commuting c-numbers. 

The non-linear sigma model is classically scale-invariant.
The function  $ {\mathbf{T}}$, on the other hand,  has naive scaling dimension of mass,  
so (classical)  scale-invariance requires that we set 
it to zero. The reader can easily check that,
in this case,  the operator (\ref{1}) is  invariant under all  conformal transformations that 
preserve $C$.  This symmetry is further enhanced if,  as a result of the field equations, 
the induced one-form
  \begin{equation} 
  \label{flat}
 {\widehat B} \equiv 
 \left(  {\mathbf{B}}_M (\Phi)    \partial_\alpha { { \Phi}}^M   +
       \epsilon_{\alpha\beta} \widetilde {\mathbf{B}}_M (\Phi) \partial^\beta \Phi^M  \right) 
       {d\zeta^\alpha}  
  \end{equation} 
is a flat $U(n)$ connection, i.e. if 
in short-hand  notation $d {\widehat B} + [ {\widehat B},  {\widehat B}] = 0$. 
The loop operator is in this case invariant under 
arbitrary continuous deformations of $C$,  as follows from the non-abelian Stoke's theorem.
Such defects can therefore be called {\it topological}. 
The eigenvalues of topological loops $W(C)$, with $C$ winding around  compact 
space,
  are charges conserved by the time evolution. The existence of a one-(spectral-) parameter
family of flat connections is, for this reason, often  tantamount to classical integrability,
see e.g.  \cite{babelon}.

  Quantization breaks, in general, the scale invariance of the defect loop even when the bulk
  theory is conformal. This is because the definition of $W(C)$ requires the introduction
  of a short-distance cutoff $\epsilon$. As the
   cutoff is being removed the couplings run 
  to infrared fixed points,  ${\mathbf{B}}^{(\epsilon)}  \to {\mathbf{B}}^{*} $ and 
  $\widetilde {\mathbf{B}}^{(\epsilon)}  \to \widetilde {\mathbf{B}}^{*} $
  as $\epsilon \to 0$.  I will
  explain  later that  this renormalization-group  flow can be described  perturbatively 
    \cite{Bachas:2004sy}
  by generalized  Dirac-Born-Infeld equations.
  The fixed-point operators commute with a diagonal conformal algebra. More specifically,
  if $C$ is the circle of a cylindrical spacetime, and $L_N$ and $\overline  L_N$ are  the left- and
  right-moving Virasoro generators, then
      \begin{equation}
  \label{Vir} 
   [ \, L_N - \overline L_{-N}\,  ,  \,  {\rm tr }_V {W}^*(C)\,  ] \, = \, 0 \ \ \ \ \ \forall N  \ .  
 \end{equation}
  There exists another class of loop operators that 
commute with the $\overline L_{N}$ (but not necessarily with the $L_N$) and which
we will call  {\it chiral}. Topological operators  lie at the intersection of the above
two classes: they  
  commute separately with the $L_N$ and  the
  $\overline  L_N$, and they are thus both conformal and chiral.

   All this  can be illustrated with the symmetry-preserving defect loops 
of the WZW model  \cite{Bachas:2004sy}. Consider the following 
  {\it chiral},  symmetry-preserving  
  defect: 
     \begin{equation}
  \label{2} 
  {\cal O}_r(C)  =  \chi_r(  \,  {\rm P} \, e^{\, i\oint_{\, C}  \, \lambda  J^a t^a}\  ) \ ,   
 \end{equation}
 where $J^a$ are the left-moving  Kac-Moody currents, $t^a$ the generators of 
 the global group $G$, and $\chi_r$ the character of the $G$-representation, $r$,   
  carried by the state-space of the defect. 
  In the classical theory  ${\cal O}_r(C) $ is  topological
   for all values of the parameter $\lambda$.  But  upon quantization,  
   the spectral parameter runs
   from the UV fixd point $\lambda^{*} = 0$ to 
   an  IR  fixed point $\lambda^{*}  \simeq 1/k$, where 
   $k$ is the level of the Kac-Moody algebra (and  $k\gg 1$ 
    for perturbation theory to be valid).  It is interesting here to note \cite{Bachas:2004sy}  that one can
    regularize (\ref{2}) while preserving the following symmetries: 
     (a) chirality, i.e. $[ {\cal O}^\epsilon_r(C),  \overline J^a_N ] = 0$ 
     for all right-moving Kac-Moody (and Virasoro) generators,
     (b) translations on the cylinder, i.e. $[ {\cal O}^\epsilon_r(C),  L_0\pm \overline L_0 ] = 0$, 
     and (c) global $G_{\rm left}$-invariance. These imply, among other things,  that  
     the RG flow can be restricted to the single parameter $\lambda$, and that the IR fixed-point
     loop operator is topological.  This fixed-point operator is the  quantum-monodromy matrix of
     the WZW model \cite{Alekseev:1990vr}. It can be constructed explicitly,  
     to all orders in the $1/k$ expansion,  as a central element
     of the enveloping algebra of the Kac-Moody  algebra \cite{Ale,Kac}. 
      
           The above renormalization-group flow describes, for $G = SU(2)$,  the screening of a magnetic impurity interacting with the left-moving spin current in a quantum wire.  This is the celebrated
 Kondo problem \cite{Wilson:1974mb}\footnote{Strictly-speaking,  in  the
 Kondo setup  the magnetic impurity
 interacts with the  s-wave conduction electrons of a 3D metal. This
 is mathematically identical to the problem  discussed here.}  which can be
 solved exactly by the   Bethe ansatz  \cite{Andrei,Wiegmann}. It was first rephrased in the
 language of conformal field theory by Affleck \cite{Affleck:1995ge}.  
Close to the spirit of our discussion here is also the work of 
Bazhanov et al \cite{Bazhanov:1994ft,Bazhanov:1996dr,Bazhanov:1998dq}, who 
proposed to study  quantum  loop operators in  minimal models 
 using conformal (as opposed to integrable lattice-model) techniques. 
 Topological loop operators  were first introduced and analyzed in CFT by Petkova 
 and Zuber \cite{Petkova:2000ip}.  Working directly in the CFT 
 makes it possible to use  the
 powerful (geometric and algebraic) tools that were  developed for the study of
 D-branes.

 \section{Interfaces as spectrum-generating symmetries}
\label{sec:2}

  Conformal defects in a sigma model  with target ${\cal M}$ can be mapped
  to conformal boundaries in a model with target ${\cal M}\otimes {\cal M}$ by the
  folding trick \cite{Oshikawa:1996dj,Bachas:2001vj}, i.e. by folding space so
  that all bulk fields live on the same side of the defect. Conformal boundaries can, 
  in turn,  be described either as  geometric D-branes \cite{Polchinski:1995mt}, or algebraically
   as  conformal boundary states on the cylinder  \cite{Callan:1987px,Polchinski:1987tu}.  
   In the latter description  space is taken to be a compact circle, and the boundary state is
   a (generally entangled) state of the two  decoupled  copies of the conformal theory:
 \begin{equation}\label{general}
\vert\hskip -0.5mm \vert \, {\cal B}\, 
\rangle\hskip -0.6mm\rangle  \ =\   \sum  {\cal B}_{\alpha_1
 \tilde \alpha_1  a_2 \tilde \alpha_2}\, 
\vert \alpha_1,\tilde \alpha_1\rangle\otimes \vert \alpha_2,\tilde \alpha_2\rangle \ . 
\end{equation}
Here $\alpha_j$ ($\tilde \alpha_j$) labels the state of the left- (right-) movers in  the
$j$th copy. Unfolding reverses the sign of time for one  copy,  
and thus  transforms the corresponding states by hermitean conjugation. 
 This converts $\vert\hskip -0.5mm \vert \, {\cal B}\, 
\rangle\hskip -0.6mm\rangle$ to a formal operator, ${\cal O}$,  which acts  on the Hilbert  space
${\cal H}$ of the conformal field theory. 
The fixed-point operators of the previous section are all, in principle,  
unfolded boundary states.

 This  discussion can be extended readily to the case where the theories
 on the left and on the right of the defect are different,  CFT$1\not=$ CFT$2$.
 Such defects should be, more properly, called {\it interfaces} or  domain walls.
 They can be  described similarly
 by a  boundary state of  CFT$1\otimes$ CFT$2$, or by the
 corresponding unfolded operator ${\cal O}_{21}
 : {\cal H}_{1} \to {\cal H}_2$. Conformal interfaces correspond to 
 operators that  intertwine the action of
 the diagonal Virasoro algebra, 
  \begin{equation}\label{diag1}
 (L_N^{(2)} - \overline L^{(2)}_{-N}) {\cal O}_{21}\,  =\,   {\cal O}_{21} (L_N^{(1)} - \overline L^{(1)}_{-N})\ ,   
  \end{equation}
while  topological interfaces intertwine separately the action of the left- and   right-movers.
  In the string-theory 
literature conformal interfaces 
were  first studied as holographic duals 
\cite{Karch:2000gx,Bachas:2001vj,DeWolfe:2001pq,Erdmenger:2002ex} 
 to  codimension-one anti-de Sitter branes \cite{Karch:2000ct,Bachas:2000fr}. 
  Note that conformal boundaries are special conformal
 interfaces for which  ${\rm CFT}2$ is the trivial theory, i.e. a theory  with
no massless degrees of freedom.   If ${\cal O}_{1\emptyset}$ is the corresponding
operator (where the empty symbol denotes the trivial theory) then 
  conformal invariance implies that
$(L_N^{(1)} - \overline L^{(1)}_{-N}){\cal O}_{1\emptyset} = 0$.

   I now  come to the main point of this  talk. Consider a closed-string background 
   described by the worldsheet theory CFT$1$, and let ${\cal O}_{1\emptyset}$ correspond to 
   a  D-brane in this background.   Take the worldsheet to be  
    the unit disk, or equivalently the semi-infinite cylinder, with the boundary described by
    the above D-brane. Consider also a conformal interface ${\cal O}_{21}$,
    where CFT$2$ describes another admissible closed-string background. 
    Now insert this  interface at infinity  and push  it to the boundary of the cylinder, as
    in figure 1.  The operation is,  in general, 
    singular except when ${\cal O}_{21}$ is a
    topological interface  in which case  it can be displaced freely.  
  
\vskip 5mm  
     
%
\begin{figure}[ht]
\centering
\hspace{40pt}
\includegraphics[scale=.40]{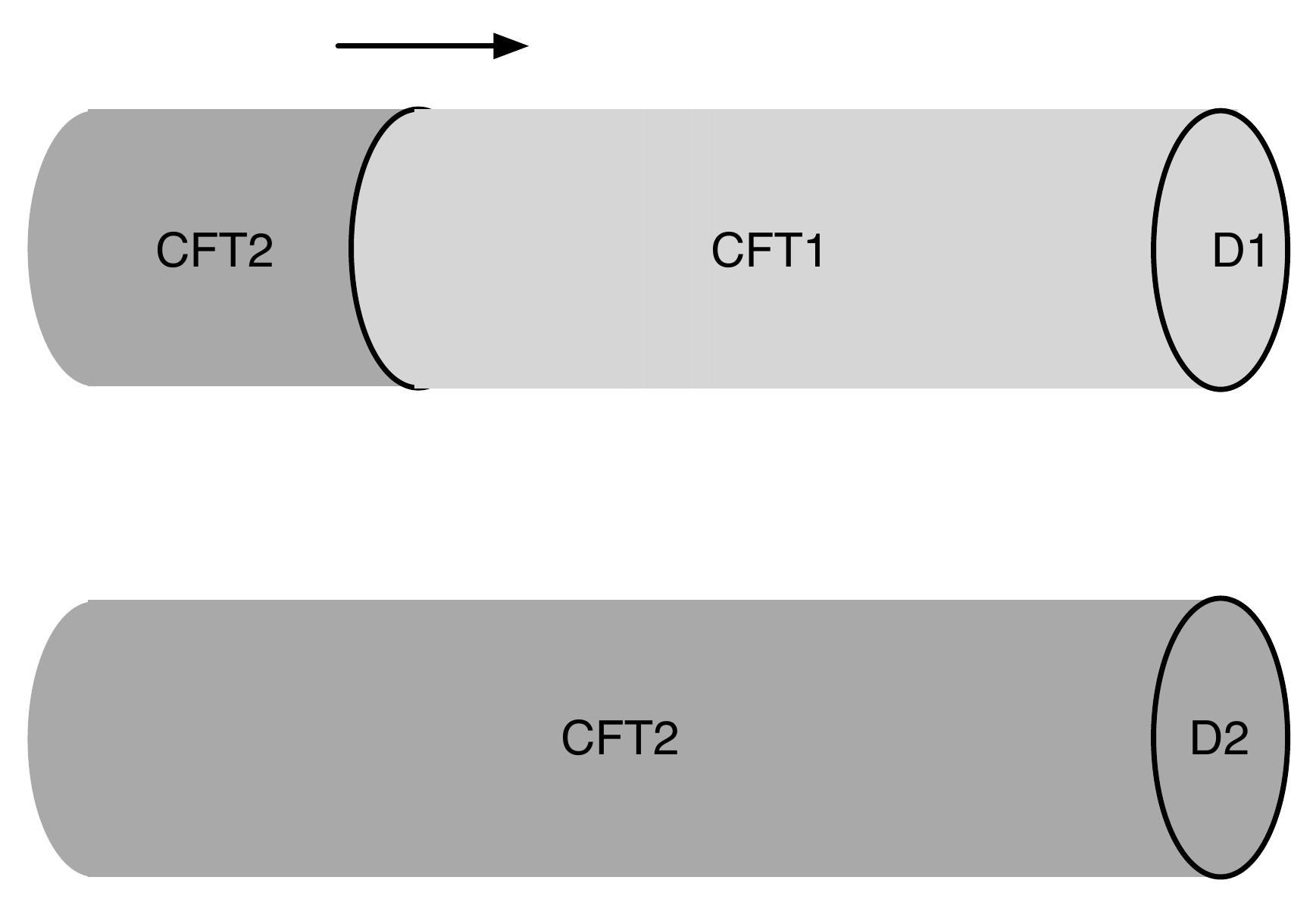}
%
%
\caption{\small An interface brought from infinity to the boundary of a
a cylindrical worldsheet  maps the D-branes of one  bulk CFT
to those of the other. Conformal interfaces between two theories  with
the same central charge act thus as spectrum-generating symmetries
of classical string theory. In many
worked-out examples these  include and
extend the perturbative dualities,  and other classical symmetries,  of the 
open- and closed-string action. 
}
\label{fig:2}       
\end{figure}
     
\vskip 5mm
  
    Let us assume,  more generally, that this fusion operation can be somehow 
    defined  and yields a  boundary state  of CFT$2$ which we denote by 
    ${\cal O}_{21}  \,  \circ  {\cal O}_{1\emptyset}$.  We assume that 
    the Virasoro generators commute past the  fusion symbol. 
    It follows then from eq. (\ref{diag1})  that the new boundary state   
      is conformal  whenever  the old one was.  Since conformal
     invariance
     is equivalent to the classical string equations, one concludes that
      ${\cal O}_{21}$ acts as a spectrum-generating symmetry of classical string theory.         
    Conformal interfaces could, in other words,  play a similar  role as the Ehlers-Geroch
     transformations  \cite{ehlers,geroch}  of General Relativity.       
       Bringing an interface to the boundary is a special case of the more general process of
       {\it fusion}, i.e.  of juxtaposing and then bringing two
       interfaces together on the string worldsheet. This  is of course only possible when the 
     CFT on the right side of  the first  interface coincides with the CFT on  the left side
      of the second.  Furthermore, two interfaces can only be added when their left and right
     CFTs are identical. Since fusion and addition  cannot  be defined for arbitrary elements, 
     the set of all conformal interfaces is neither an algebra nor a
     group.
     By abuse of language, I will nevertheless refer to it
     as the ``interface algebra".\footnote{The correct term for the
      interfaces is ``functors". For a more  accurate
     mathematical terminology   the reader should consult,
       for instance,  reference  \cite{Fuchs:2007vk}.}

     The first thing to note is that the interface
       ``algebra"   is  non-trivial even if restricted  only to  
       elements  with non-singular fusion.  These include  all the topological interfaces,   
     for which  fusion is  the
      regular  product of the corresponding operators, 
      ${\cal O}_A\circ {\cal O}_B= {\cal O}_A {\cal O}_B$. 
    The simplest topological defects
    are those whose internal state is decoupled from the dynamics in the bulk. 
    They   correspond to multiples of the identity operator, ${\cal O} = n {\bf 1}$ with $n$ a natural
    number. Their action on  any D-brane endows this latter  with Chan-Paton multiplicity. 
    Less trivial are the topological defects which generate symmetries of the CFT, as well as
    the topological interfaces that generate  perturbative T-dualities. These were first studied,
    for several  examples, 
    in two beautiful papers by Fr\"ohlich et al \cite{Frohlich:2004ef, Frohlich:2006ch}.  
    The fact that all  perturbative string symmetries can be realized through the action of
    local defects is not a priori obvious (and needs still to be generally established). 
 Other interesting examples are the minimal-model topological defects,   
 shown to generate universal boundary flows  
    \cite{Graham:2003nc,Fredenhagen:2003xf}. A different
    set of conformal interfaces whose fusion is non-singular are those that preserve 
    at least $N=(2,2)$ supersymmetry  \cite{Brunner:2007qu,Brunner:2007ur}. 
    Some of these descend from  supersymmetric gauge theories in higher dimensions
    \cite{Kapustin:2006pk,Gukov:2006jk,Kapustin:2006hi,Kapustin:2007wm}.
   Such interfaces were,  in particular,  used to  generate the monodromy
    transformations of supersymmetric  D-branes transported around singular points
    in the Calabi-Yau 
    moduli space  \cite{Brunner:2008fa}.  As these and other examples demonstrate,    
    the interface ``algebra"  is  very rich  even if restricted
  to  interfaces with non-singular fusion.

       Extending the  structure to arbitrary interfaces is, nevertheless, an   interesting problem. 
Firstly, the algebras (without quotation marks)  of non-topological defects
would provide, if they could be defined,  large extensions 
of the automorphism groups of various CFTs. 
Furthermore,  while topological interfaces are rare -- they may only join  CFTs  that have
    isomorphic  Virasoro representations  --  the conformal ones are on the contrary  common.
      To see that conformal interfaces are not rare, consider  the $n$th multiple of the
     identity defect which is mapped,   
   after folding,   to  $n$   diagonally-embedded 
   middle-dimensional branes in ${\cal M}\times {\cal M}$
    \cite{Bachas:2001vj}. A generic  Hamiltonian of the form (\ref{gen1form}),  with
    the tachyon potential ${\bf T}$ set to zero, corresponds  to 
    arbitrary geometric and gauge-field perturbations
    of  these diagonal branes. Any solution of the (non-abelian, $\alpha^\prime$
    corrected) Dirac-Born-Infeld equations for  these branes  gives therefore rise
    to a conformal defect \cite{Bachas:2004sy}.  Likewise, any non-factorizable D-brane
    of  CFT$1\otimes$CFT$2$ unfolds to a non-trivial  
     interface between the two conformal field theories.    All of these interfaces can be
     characterized by a  reflection coefficient,  
    ${\cal R}$,  \cite{Quella:2006de}  which must vanish in the topological case.

     For most of these interfaces the products of the corresponding operators are  singular,
     so the fusion needs to be appropriately defined.  
       A first step in this direction was taken, in the context of a
       free-scalar  theory, in reference  \cite{Bachas:2007td}. 
       The rough idea is to define the fusion product as the renormalization-group 
      fixed point to which the system of the two interfaces  flows  when their  separation, $\epsilon$, 
      goes to zero.  A systematic  way of doing this, consistent 
      with the distributive property of fusion,\footnote{I thank Maxim Kontsevich for stressing this point.} 
       has not yet been worked out for interacting theories.  For free fields, on the other hand, 
  the story is simpler.  The  short-distance singularities  are in this case 
  expected to be of the general  form 
      \begin{equation}\label{product1}
   {\cal O}_{A}  \,  e^{-\varepsilon(L_0  + \overline L_0)}\,     {\cal O}_{B}  \   
   \simeq \      \sum_C \,  (e^{2\pi/\varepsilon}) ^{d_{AB}^{\ C}}\  N_{AB}^{\ C}\,   {\cal O}_{C} \ ,   
  \end{equation}
where  $\epsilon\simeq 0$ is the separation of the two (circular)  interfaces
on the cylinder, $L_0  + \overline L_0$ is the
translation operator in  the middle CFT,  
  the $d_{AB}^{\ C}$ are (non-universal)
constants,   and the $N_{AB}^{\ C}$ are integer multiplicities. 
The singular coefficients in the above expression are Boltzmann factors
for divergent Casimir energies. The latter must be proportional to $1/\varepsilon$ 
which is  the only
 scale in the problem (other than 
  the inverse temperature normalized to   $\beta = 2\pi$).

    By analogy with the operator-product expansion and the
 Verlinde algebra \cite{Verlinde:1988sn}  
 we may extract  from expression (\ref{product1}) the fusion rule
         \begin{equation}\label{product}
   {\cal O}_{A} \circ  {\cal O}_{B} \  =  \  \sum_C\,    N_{AB}^{\ C}\,   {\cal O}_{C} \ . 
     \end{equation}
The following  iterative
argument shows that this definition respects the conformal symmetry: 
first multiply the left-hand-side of (\ref{product1})
with the most singular inverse Boltzmann factor (the one with the largest ${d_{AB}^{\ C}}$) and
 take  the limit $\varepsilon \to 0$ so as to extract  the leading term of the product. 
Since  $[ L_N - \overline L_{-N}, e^{-\varepsilon(L_0  + \overline L_0)} ] \simeq o(\varepsilon)$  
the result commutes with the diagonal Virasoro algebra.  Next subtract the  leading  term
from the left-hand-side of  (\ref{product1}), 
 and mutliply by the inverse Boltzmann factor with the second-largest ${d_{AB}^{\ C}}$. 
 This picks up the subleading term which, thanks to the above argument and the conformal
 symmetry of the leading term,  
  commutes also  with the diagonal Virasoro algebra. Continuing this iterative
 reasoning  proves that the right-hand-side of (\ref{product})  is  conformal as claimed.

  \section{The $c=1$ CFT  and a black hole analogy}
\label{sec:3}  

      A simple context in which to  illustrate the above ideas is the $c=1$ conformal theory
      of a periodically-identified free scalar field, $\phi  = \phi + 2\pi R$.  
      Consider the interfaces that preserve a  $U(1)\times U(1)$ symmetry, i.e. those
      described by  linear gluing conditions for the field $\phi$.  
     They correspond,  after folding,  to combinations of 
      D1-branes and of magnetized D2-branes  on the orthogonal two-torus  whose
      radii,  $R_1$ and $R_2$,  are the radii on either side of the interface. 
       The  D1-branes
      are characterized by their  winding numbers,  $k_1$ and $k_2$, and
      by the  Wilson line  and periodic position 
      moduli $\alpha$ and $\beta$.   The  magnetized D2-branes are obtained 
      from the D1-branes by T-dualizing one of the two directions of the torus -- they have therefore
      the same number of discrete and of continuous  moduli. 
      
          Let us focus here on the D1-branes. The   corresponding boundary states read
    \begin{equation}\label{k1k2}
 \vert\hskip -0.5mm \vert \, {\rm D}1,  \vartheta \rangle \hskip-0.6mm \rangle =  {g}^{(+)}  
\prod_{n=1}^\infty   (e^{S^{(+)}_{ij} a_n^i \widetilde a_n^j})^\dagger \ 
  \sum_{N,M =-\infty}^\infty  
e^{i{N} \alpha -iM \beta}  \vert k_2N, k_1M
\rangle \otimes \vert
- k_1N,  k_2M  \rangle\ ,  
\end{equation}
where $a_n^j$ and $\tilde a_n^j$ are the left- and right-moving annihilation
operators of the field $\phi_j$ (for $j=1,2$) and the dagger denotes hermitean
conjugation. The ground states 
$\vert m, \tilde m
\rangle $   of the scalar fields are characterized by
a momentum ($m$)  and a winding number ($\tilde m$). The states
in the above tensor product  correspond  to 
 $\phi_1$  and $\phi_2$.  Furthermore
 \begin{equation}\label{Sold}
 S^{(+)}  =  {\cal U}^T(\vartheta)
  \left(\hskip -1mm \begin{array}{cc}  -1 & 0 \\
                              0 & 1  \end{array}\hskip -1mm \right) \hskip -0.6mm
                               {\cal U}(\vartheta)
                               =
                                \left(\hskip -1mm \begin{array}{cc}
  -{\rm cos\, 2\vartheta} &  -{\rm sin\, 2\vartheta} \\
                               - {\rm sin\, 2\vartheta} & {\rm cos\,
                                 2\vartheta}
  \end{array} \hskip -1mm\right)  \ ,                            
 \end{equation}
where ${\cal U}(\vartheta)$ is a rotation matrix and
$\vartheta = {\rm arctan} (k_2R_2/k_1R_1)$ is  the angle between the D1-brane
and the $\phi_1$ direction.  
Finally,   the normalization constant is    
the $g$-factor  \cite{Affleck:1991tk} of the boundary state. It  is given by
 \begin{equation} 
\label{gplus}
{g}^{(+)} \, =\, {\ell\over \sqrt{2V}} \, = \,  \sqrt{k_1^2R_1^2+k_2^2 
R_2^2\over  2R_1R_2}\,  = \,   \sqrt{
 k_1 k_2  \over {\rm sin} 2\vartheta}\ ,  
 \end{equation} 
where  $\ell$ is  the length of the D1-brane,   $V$  the volume of the two-torus, 
and the last  rewriting  
 follows from straightforward trigonometry.   The logarithm of the  
 $g$ factor  is the invariant entropy of the interface. 
 
  Inspection of the expression  (\ref{k1k2})  shows that
  the non-zero modes of the fields $\phi_j$ are only sensitive to  
 the angle $\vartheta$,  which also determines the reflection 
 coefficient  of the interface \cite{Quella:2006de}.   For fixed $k_1$ and $k_2$ the 
  $g$ factor is minimal   when $\vartheta = \pm \pi/4$, in which
  case the reflection  ${\cal R}=0$ and
   the interface is  topological. 
Note that this requirement fixes the ratio of the two bulk moduli: $R_1/R_2 = \vert k_2/k_1\vert $. 
When $\vert k_1\vert$ = $\vert k_2\vert$ = 1 the two radii are equal and the 
invariant entropy is zero.
The corresponding  topological defects generate the automorphisms of the CFT, i.e. 
sign flip  of the field $\phi$ and separate
 translations  of its left- and right-moving pieces. 
The identity defect corresponds to the diagonal D1-brane, 
with   $k_1= k_2 = 1$ and $\alpha = \beta = 0$. 
A T-duality along $\phi_1$ maps this 
topological defect to a D2-brane with one unit of magnetic flux. The
corresponding interface operator is  the generator of the radius-inverting  T-duality transformation. 
 All other topological interfaces  have
 positive entropy,  ${\rm log}g = {\rm log}\sqrt{\vert k_1k_2\vert} > 0 $. 
 One may conjecture that the following statement  is more generally true \cite{Bachas:2007td}: the  entropy
of all topological interfaces is non-negative, and it vanishes only  for 
  CFT automorphisms.   
 
         The interfaces given by equations (\ref{k1k2}) to (\ref{gplus})
         exist for all values of the bulk radii $R_1$ and $R_2$. By
  choosing the radii  to be equal we obtain a large set of conformal defects whose algebra is an  
  extension of the automorphism group of the CFT. 
  For a  detailed derivation of this algebra see reference \cite{Bachas:2007td}.  
  The  fusion  rule for the discrete defect moduli turns out to be multiplicative, 
 \begin{equation}  
  \label{equivc1}
[ k_1, k_2; s] \circ [ k_1^\prime, k_2^\prime; s^\prime] \ 
=\ [ k_1k_1^\prime, k_2k_2^\prime; s s^\prime] \ , \nonumber 
 \end{equation}  
where  $[ k_1, k_2; s]$ denotes a defect with integer moduli $k_1$, $k_2$, $s$, where 
$s = +, -\ $ according to whether the folded defect is a D1-brane or a 
magnetized D2-brane.  The above fusion  rule continues to hold for general interfaces,
i.e. when  the radii on either side are not the same.  Let me also give the composition
rule for the angle $\vartheta$ in this general case (assuming 
  $s=s^\prime=+$):  
 \begin{equation} \label{equivc}
{\rm tan}(\vartheta\circ\vartheta^\prime) = {\rm tan} \vartheta\,  {\rm tan}\vartheta^\prime\ , 
 \end{equation}
where $\vartheta\circ\vartheta^\prime$ denotes the angle of the fusion product. 
The composition rule (\ref{equivc1}) was
first derived, for the  topological  interfaces, in reference \cite{Fuchs:2007tx}. 
 In this case the tangents in the last equation are $\pm 1$ and all operator products
 are non-singular. 

      There exist  some intriguing similarities \cite{Bachas:2007td}
between the above conformal interfaces and supergravity black holes.  
 The counterpart of  BPS black holes are the topological interfaces, 
 which  (a) minimize the free energy for fixed
 values of the discrete charges, (b) fix through an ``attractor mechanism"  \cite{attractor}
 a combination
 of the bulk moduli, and (c) are marginally stable against dissociation -- the inverse
 process of fusion.  The interface ``algebra"  is,  in this sense,  reminiscent of  an earlier
 effort by Harvey and Moore \cite{HaMo} 
  to define an extended symmetry algebra for string theory.
 Their symmetry generators were vertex operators for supersymmetric states
 of the compactified  string. One noteworthy difference is that  the
 additively-conserved  charges in the above example are logarithms of natural numbers, 
  rather than taking values in a charge lattice as in \cite{HaMo}.     
 Whether these observations have any deeper meaning remains to be seen.
 Another direction
  worth exploring is a possible relation of the above ideas with efforts to formulate string
  theory in a ``doubled geometry",  see for instance \cite{Hull:2006va}.
 The doubling of spacetime after folding suggests that 
 such a formalism may be the natural language in which  the defect algebras should be 
 formulated and discussed.

 Time now to conclude:  conformal interfaces and defects are  examples of extended
 operators,  which are a rich and still only partially-explored chapter of quantum field
 theory.  They describe a variety of critical phenomena in low-dimensional 
 condensed-matter systems which, for lack of time, I have not discussed.
Conformal interfaces can be added and, at  least in many studied examples,  juxtaposed or fused. 
The resulting  interface ``algebra" defines a large extension of the classical 
string symmetries, which deserves  to be studied more.  
 
 \vskip 5mm

 {\bf Acknowledgements} \hfil\break
I   thank Ilka Brunner,  J\"urg Fr\"ohlich and Samuel Monnier for
very pleasant collaborations during the last couple of years,  on different aspects of
this talk. Many thanks also to
Eric D'Hoker, Mike Douglas, Sergei Gukov, Chris Hull  and Maxim Kontsevich for useful conversations
and comments.
Claudio Bunster supervised my senior-year undergraduate  thesis,  and helped me publish my first 
scientific article --  always a source of considerable pride for a student.
My gratitude, after all these years, remains intact. 
 This work
has been supported in part by the European Community Human
Potential Program under contracts MRTN-CT-2004-005104 and
MRTN-CT-2004-512194.

\end{document}